\documentclass[a4paper,12pt,leqno]{article}
\usepackage{theorem}
\usepackage{latexsym,amssymb,amsfonts,amsmath}
\usepackage[dvips]{graphicx}
\newtheorem{theorem}{Theorem}
\newtheorem{lemma}{Lemma}
\newtheorem{corollary}[theorem]{Corollary}

\newtheorem{remark}[theorem]{Remark}

\newcommand{\AS}{{\bf Assumption} }
\newcommand{\As}{{\bf Assumption} }
\newcommand{\IND}{{\chi}}
\newcommand{\ID}{{\bf 1}}
\newcommand{\bs}[1]{\boldsymbol{#1}}
\newcommand{\DTQW}[1]{X^{(D)}_{#1}}
\newcommand{\CTQW}[1]{X^{(C)}_{#1}}

\newcommand{\DTQWPM}[1]{\widetilde{X}^{(D)}_{#1}}
\newcommand{\CTQWPM}[1]{\widetilde{X}^{(C)}_{#1}}
\newcommand{\PPM}[2]{Y^{(#2)}_{{\sf{#1}}}}
\newcommand{\FTD}[1]{X^{(F)}_{#1}}
\newcommand{\QW}[1]{W^{(F)}_{#1}}
\newcommand{\LA}[1]{Z^{(F)}_{#1}}
\newcommand{\CO}[1]{\widetilde{Z}^{(F)}_{#1}}

\title{{\Large {\bf CROSSOVERS INDUCED BY DISCRETE-TIME QUANTUM WALKS 
}
}}

\author{ 
{\small 
KOTA CHISAKI,$^{1}$ \footnote{chisaki@npde.osu.sci.ynu.ac.jp}\quad 
NORIO KONNO,$^{2}$ \footnote{konno@ynu.ac.jp} \quad 
ETSUO SEGAWA,$^{3}$ \footnote{To whom correspondence should be addressed. segawa@stat.t.u-tokyo.ac.jp}\quad
YUTAKA SHIKANO$^{4}$ \footnote{shikano@th.phys.titech.ac.jp} 
}\\
{\scriptsize $^{1,2}$ 
Department of Applied Mathematics, Faculty of Engineering, Yokohama National University
}\\
{\scriptsize Hodogaya, Yokohama 240-8501, Japan
} \\
{\scriptsize $^3$ 
Department of Mathematical Informatics, The University of Tokyo,
}\\
{\scriptsize Bunkyo, Tokyo, 113-8656, Japan
} \\
{\scriptsize $^4$ 
Department of Physics, Tokyo Institute of Technology,
}\\
{\scriptsize  
 Department of Mathematical Engineering, Massachusetts Institute of Technology,
}\\
{\scriptsize Meguro, Tokyo 152-8551, Japan
} \\
} 

\date{\empty }
\begin{document}
\maketitle

\par\noindent
\begin{small}
\par\noindent
{\bf Abstract}. We consider crossovers with respect to the weak convergence theorems from a discrete-time quantum walk (DTQW). 
We show that a continuous-time quantum walk (CTQW) and 
discrete- and continuous-time random walks can be expressed as 
DTQWs in some limits. 
At first we generalize our previous study 
[Phys. Rev. A \textbf{81}, 062129 (2010)] on the DTQW with position measurements. 
We show that the position measurements per each step 
with probability $p \sim 1/n^\beta$ can be evaluated, where $n$ is the final time and $0<\beta<1$. 
We also give a corresponding continuous-time case. 
As a consequence, crossovers from the diffusive spreading (random walk) to the ballistic spreading (quantum walk) 
can be seen as the parameter $\beta$ shifts from 0 to 1 in both discrete- and continuous-time cases of 
the weak convergence theorems. 
Secondly, we introduce a new class of the DTQW, in which the absolute value of the diagonal parts of the quantum coin 
is proportional to a power of the inverse of the final time $n$. This is called a final-time-dependent DTQW (FTD-DTQW).
The CTQW is obtained in a limit of the FTD-DTQW. 
We also obtain the weak convergence theorem for the FTD-DTQW which shows a variety of spreading properties. 
Finally, we consider the FTD-DTQW with periodic position measurements. 
This weak convergence theorem gives a phase diagram which maps sufficiently long-time behaviors of the 
discrete- and continuous-time quantum and random walks.

\footnote[0]{
{\it Key words and phrases.} 
Quantum walk, crossover and weak convergence. 
}

\end{small}

\setcounter{equation}{0}
\section{Introduction}
\quad A quantum walk (QW) is a quantum analogue of the random walk (RW)~\cite{Gudder,Aharonov,Meyer}. 
As is the RW has important roles in various fields, 
it has been shown that the QW also plays important roles in the quantum world, for example, 
constructing quantum speed-up algorithm~\cite{Ambainis} and expressing the energy transfer on 
the chromatographic network in the photosynthetic system~\cite{Alan}. 
It is shown that approximations of a discrete-time quantum walk (DTQW) give 
the Dirac equation or a continuous-time quantum walk (CTQW), \textit{i.e.}, 
the discretized Schr\"{o}dinger equation~\cite{Strauch3,Strauch2,Strauch1,Chandra}.
The behaviors of the RW and the corresponding QW are quite different. 
One of the effective tools to show this difference is the weak convergence theorem~\cite{KonnoQL0,KonnoQL1}. 
For simplicity, we restrict the quantum and random walks 
on the infinite one-dimensional lattice $\mathbb{Z}$ throughout this paper. 
The diffusive spreadings of a symmetric discrete-time random walk (DTRW) and a continuous-time 
random walk (CTRW) 
follow from the central limit theorems. 
On the other hand, 
as shown in Refs.~\cite{KonnoQL0,KonnoQL1,KonnoQLC}, the weak convergence theorems 
for the corresponding DTQW and CTQW, give 
ballistic spreadings of the QWs with inverted bell shaped limit densities. 

The power of the time variable in the weak convergence for the QW doubles RW's one. 
As is seen in Refs.~\cite{Brun,Zhang,Kendon} for example, 
it is shown that random or periodic measurements of a DTQW induces a sudden 
\textit{transition} from the quantum to the random walk. 
In our previous study~\cite{SCSK}, 
we introduced a toy model as an approximation of the DTQW with periodic position measurements. 
We showed a gradual \textit{crossover} from the DTQW to the DTRW 
with respect to the weak convergence when 
the number of position measurements in the walk is given by a power of the final time. 
The state of the DTQW treated here is described by a direct product of position 
and coin states \cite{Meyer,Ambainis}. 
In the DTQW with position measurements (DTQW with PM), the following procedure are repeated: \\

\noindent 
{\it Procedure 1.} (DTQW with PM) \\
Let an sequence be $(t_1,t_2,\dots,t_M)$ with $t_j=\sum_{i=1}^{j}d_i$, 
where $d_i$'s are given by i.i.d. geometric distribution with a parameter $p$. 
We call $t_j$ ($j$-th) measurement time. 
\begin{enumerate}
\item Measure the position state of the walk at each measurement time. 
\item Restart the DTQW from the measurement position with the normalized coin state of the position. 
\item Keep the DTQW by the next measurement time. 
\item Repeat (1)-(3) until the final time $t_M$ comes.
\end{enumerate}
On the other hand, we studied in Ref. \cite{SCSK} on the QW given by the following Procedure 2 
and gave a crossover from ballistic to diffusive behaviors: \\

\noindent
{\it Procedure 2.} (Periodic position measurements~\cite{SCSK}) \\
Let the sequence of measurement times be $(d,2d,\dots,Md)$ with $n=Md$, $\lim_{n\to\infty}d/n^{\beta}=1$ $(\beta\in [0,1])$. 
Prepare a {\it fixed} coin state $\bs{\varphi}$ with $||\bs{\varphi}||=1$. 
\begin{enumerate}
\item Measure the position state of the walk at each measurement time. 
\item Restart the DTQW from the measurement position with the coin state $\bs{\varphi}$. 
\item Keep the DTQW by the next measurement time. 
\item Repeat (1)-(3) until the final time $n$ comes. 
\end{enumerate}
In general, the spans between the measurements and the reprepared initial coin states 
in Procedure 1 are varied unlike the walk treated in Ref. \cite{SCSK}. 
In this paper, we improve the walk of Ref. \cite{SCSK} as follows: \\

\noindent 
{\it Procedure 3}. (The walk treated in this paper)\\
Let the sequence of measurement times be $(s_1,s_2,\dots,s_M)$ with $s_1<s_2<\cdots<s_M$. 
Put a \textit{prepared} sequence of coin states by $(\bs{\varphi}_1,\bs{\varphi}_2,\dots,\bs{\varphi}_M)$ with $||\bs{\varphi}_j||=1$ $(j=1,2,\dots,M)$. 
\begin{enumerate}
\item Measure the position state of the walk at each measurement time $s_j$. 
\item Restart the DTQW from the measurement position with the normalized coin state $\bs{\varphi}_j$. 
\item Keep the DTQW by the next measurement time $s_{j+1}$. 
\item Repeat (1)-(3) updating the subscript $j$ until the final time $s_M$ comes.
\end{enumerate}
%
In Procedure 3, the measurement times can be taken as a random variable while the coin state after each measurement is {\it a priori} given like Procedure 2. 
It is remarked that 
the diffusive spreading of the DTQW with both \textit{position} and \textit{coin} measurements 
with the parameter of the geometric distribution $p$ in the weak convergence theorem was shown 
in Ref. \cite{Zhang}. 
On the other hand, we find a crossover from the diffusive to the ballistic spreading 
through the sub-ballistic (super-diffusive) spreading in some class of the walks of Procedure 3 
including the walk in Ref. \cite{SCSK} from the weak convergence theorem. 
We also give a similar result on a crossover of a corresponding continuous-time case. 

The similarity of DTQWs and the corresponding CTQWs can be seen in various QW's models 
on $\mathbb{Z}$~\cite{KonnoQL0,KonnoQLC} and homogeneous trees~\cite{CHKS,KonnoT}. 
Furthermore, the relationship is shown on $\mathbb{Z}$ in Ref. \cite{Strauch1} 
and general graphs in Refs. \cite{DAlessandro,Childs}. 
To discuss on the similarity between DTQW and CTQW with PM, 
in this paper, we 
introduce a final-time-dependent DTQW (FTD-DTQW) characterized by the quantum coin 
whose absolute value of diagonal parts is in inverse proportion to a power of the final time. 
This is a modified version of Refs. \cite{Strauch1,Romanelli}. 
In Ref. \cite{Strauch1}, the absolute value of diagonal parts of the quantum coin is given by sufficiently 
small $\epsilon$ which is independent of each time step. 
It is shown that an asymptotic expansion of the DTQW with respect to small $\epsilon$ gives the corresponding CTQW. 
Also in Ref. \cite{Romanelli}, the absolute value of diagonal parts of 
the time dependent quantum coin is in inverse proportion to a power of each time. 
Various spreading properties of this walk, 
\textit{i.e.}, ballistic, sub-ballistic, diffusive, sub-diffusive, and localized, are shown. 
We show that a limit of the FTD-DTQW gives the corresponding CTQW and a relation 
between the discrete final time $n$ and the corresponding continuous time $t$: 
the ratio $n/t$ is an important value to give the weak convergence theorems associated with 
the spreading properties. 
We find that both the discrete- and continuous-time QWs with position measurements are 
described as special cases of the FTD-DTQW in the long time limit. 

The remainder of this paper is organized as follows. 
In Sect. \ref{sec2}, we give definitions of the discrete- and continuous-time QWs. 
Section \ref{sec3} treats the walk following Procedure 3, which is an extended model of Ref.~\cite{SCSK}. 
In the walk following Procedure 2, we showed in Ref.~\cite{SCSK} that 
a crossover from the DTQW to the DTRW with periodic position measurements (PPM) appeared. 
We find that even if the measurements are not periodic, and re-preparing initial coin states 
are varied, the crossover can be also seen in some conditions. 
Furthermore, we also give a similar result on the crossover of the CTQW corresponding to the DTQW~\cite{SCSK}. 
Section \ref{sec4} presents an FTD-DTQW. 
As shown in Ref.~\cite{Strauch1}, we give a CTQW corresponding to the FTD-DTQW. 
We see a crossover from the ballistic to the localized spreading of the FTD-DTQW 
in the weak convergence theorem. 
To clarify the relation between the crossovers in the discrete- and continuous-time models, 
we consider a hybrid type of the walks, \textit{i.e.}, the FTD-DTQW with PPM in Sect. \ref{sec5}. 
Our analytical method is based on the Fourier analysis~\cite{Grimmett,SK}. 
As a consequence, the weak convergence theorem for the FTD-DTQW with PPM shows that 
the DTQW is one of the fundamental processes which give asymptotic behavior of the corresponding CTQW 
and CTRW and DTRW. Section \ref{sec6} is devoted to summary. 
\section {Definition of discrete- and continuous-time QWs on $\mathbb{Z}$} \label{sec2}
\begin{enumerate}
\item{DTQW:} 
The state space of the one dimensional DTQW with a two dimensional coin is defined as a tensor product of the position-state space $\mathcal {H}_P$ and 
the coin-state space $\mathcal{H}_C$, where 
$\mathcal{H}_P$ and $\mathcal{H}_C$ are associated with the orthogonal bases $\{ \boldsymbol{\delta}_x: x\in \mathbb{Z}\}$ and 
$\{\bs{e}_L,\bs{e}_R\}$, respectively. 
The one-step unitary time evolution for the DTQW is given by $U=S(\ID_P\otimes H)$, where 
$S$ is a shift operator such that 
$S (\bs{\delta}_x \otimes \bs{e}_R)=\bs{\delta}_{x+1}\otimes \bs{e}_R$ and 
$S (\bs{\delta}_x \otimes \bs{e}_L)=\bs{\delta}_{x-1}\otimes \bs{e}_L$, 
$\ID_P$ is the identity operator on $\mathcal{H}_P$, and 
$H$ is a 2-dimensional unitary operator called ``quantum coin'' on $\mathcal{H}_C$ described by 
\begin{equation}
H=\begin{bmatrix}a & b \\ c & d \end{bmatrix} 
\end{equation}
with $abcd \neq 0$. 
Here $\bs{e}_{L} \equiv [1,0]^\dagger$ 
and $\bs{e}_{R} \equiv [0,1]^\dagger$ throughout this paper, 
where $A^\dagger$ means the conjugate and transpose of $A$. 
Put $\bs{\Psi}_n^{(D)}(x)$ as the coin state at time $n$ and position $x$. 
Then the coin state $\bs{\Psi}_n^{(D)}(x)$ has the following recurrence relation, that is, 
\begin{align} 
	\bs{\Psi}_n^{(D)}(x) &= Q\bs{\Psi}_{n-1}^{(D)}(x-1)+P\bs{\Psi}_{n-1}^{(D)}(x+1) \;\; (n\geq 1), \notag \\ 
	\bs{\Psi}_0^{(D)}(x) &= \delta_0(x) \bs{\varphi}_0, 
\end{align}
where $P=\bs{e}_L\bs{e}_L^{\dagger}H$, $Q=\bs{e}_R\bs{e}_R^{\dagger}H$, $\delta_0(x)$ is the Dirac delta function of $x$, and 
$\bs{\varphi}_0$ is the initial coin state:  
$\bs{\varphi}_0=q_L\bs{e}_L+q_R\bs{e}_R$ with $|q_L|^2+|q_R|^2=1$. 
Let $X^{(D)}_n$ be 
the DTQW at time $n$. 
The distribution of $\DTQW{n}$ is defined by 
$\mu_n(x)\equiv P(\DTQW{n}=x)=||\bs{\Psi}_n^{(D)}(x)||^2$. 
Note that by the unitarity of the time evolution $U$ and the normality of the initial state, 
the distribution sequence $(\mu_0,\mu_1,\mu_2,\dots)$ for each time is defined. 
It is shown that if an initial coin state $\bs{\varphi}^{(s)}=q_L^{(s)}\bs{e}_L+q_R^{(s)}\bs{e}_R$ satisfies the symmetric condition~\cite{KonnoQL0,KonnoQL1},
\begin{equation}\label{symmetric}
|q_R^{(s)}|=|q_L^{(s)}|=1/\sqrt{2}, \ \ aq_L^{(s)}\overline{bq_R^{(s)}}+\overline{aq_L^{(s)}} bq_R^{(s)}=0, 
\end{equation}
where $\overline{c}$ is the complex conjugate to $c\in \mathbb{C}$, 
then $\DTQW{n}/n$ converges weakly to $K(a)$ whose density is explicitly expressed as~\cite{KonnoQL0,KonnoQL1}
\begin{equation}\label{konnodens}
\rho_K(x;a) = \frac{\sqrt{1-|a|^2}}{\pi (1-x^2)\sqrt{|a|^2-x^2}} \IND_{(-|a|,|a|)}(x), 
\end{equation}
where, $\IND_{\Omega} (x)$ is the indicator function of the region $\Omega$ of $x$.
Remark that $\rho_K$ 
depends only on the norm of the left top element of the quantum coin defined in Eq. (1). 
\item{CTQW:}
The state space is defined as the position state space $\mathcal{H}_P$ only in contrast with the DTQW. 
Let $\Psi_t^{(C)}(x)$ be the state at time $t$ at position $x$. The time evolution is given by 
the discretized Schr\"{o}dinger equation: 
\begin{align}\label{defConti}
-i\frac{\partial \Psi^{(C)}_t(x)}{\partial t}
	&=\frac{1}{2}\left(\gamma \Psi^{(C)}_t(x-1)+\overline{\gamma}\Psi^{(C)}_t(x+1)\right)\;\;(t>0), \notag \\
\Psi_0^{(C)}(x) &= \delta_0(x), 
\end{align}
where $\gamma$ is a complex number. 
Let $\CTQW{t}$ be the CTQW at time $t$. 
The distribution of $\CTQW{t}$ is given by $P(\CTQW{t}=x)=|\Psi^{(C)}_{t}(x)|^2$. 
It is shown in Ref. ~\cite{KonnoQLC} that $\CTQW{t}/t$ converges weakly to $Z(\gamma)$ whose density 
corresponds to a scaled arcsine law~\cite{KonnoQLC}:  
\begin{equation} \label{arcsine}
\rho_Z(x;\gamma) = \frac{\IND_{(-|\gamma|,|\gamma|)}(x)}{\pi\sqrt{|\gamma|^2-x^2}}. 
\end{equation}

\end{enumerate}
\section{Crossover from QW to RW} \label{sec3}
To give the relation between RW and QW, we consider the QW with position measurements in the following. 
Let $(d_1(s),d_2(s),\dots)$ be a sequence of functions of a parameter $s$, where 
in the case of discrete (resp. continuous) time, $s$ is a natural (resp. non-negative real) number. 
The value $d_j(s)$ corresponds to the span between $(j-1)$-th and $j$-th measurements. 
According to Ref. \cite{CHKS}, we take $s^\beta$ as $d_j(s)$ for all $j\in \{1,2,\dots\}$ with $0\leq \beta \leq 1$. 
In this paper, we consider the more general setting by imposing $d_j(s)$ under the following assumption: \\
\noindent \\
{\bf Assumption}: 
For all $j\in \{1,2,\dots\}$,
\begin{enumerate}
\item $d_j(s)\leq s$, \label{as1} 
\item $d_j(s)\to \infty \ {\rm as} \ s \to \infty$, \label{as2}
\item $d_j^2(s) / \{\sum_{l=1}^{M(s)}d_l^2(s)\} \to 0 \ {\rm as} \ s \to \infty$, \label{as3}
\end{enumerate}
where $M(s) \equiv \mathrm{sup}\{m: \sum_{j=1}^m d_j(s)\leq s\}$. \\

From \AS (\ref{as2}) and (\ref{as3}), both $d_j(s)$ and $M(s)$ go to infinity as $s\to\infty$ simultaneously. 
We measure the position state of a QW 
at times $d_1(s)$, $d_1(s)+d_2(s)$,...,$d_1(s)+d_2(s)+\cdots+d_{M(s)}(s)$, and restart the QW 
by the next measurement time after each measurement. 
Thus $M(s)$ corresponds to the number of measurements. 
It should be noted that 
there is no difference of the behavior between the RW with and without position measurements. 
However, we will see that the both DTQW and CTQW give essential changes in the weak convergence theorems 
by the position measurements in the following subsections. 
\subsection{DTQW with position measurements} \label{sec3:1}
Let $\PPM{j}{D}$ be the DTQW at time $d_j(n)$ with a quantum coin $H$ 
and with an initial coin state $\bs{\varphi}_j$ $(j=1,2,\dots )$. 
We consider a convolution of an independent sequence $\{\PPM{j}{D}\}_{{\sf j}=1}^{M(n)}$ denoted as 
$\DTQWPM{n} \equiv \PPM{1}{D} + \PPM{2}{D} +\cdots + \PPM{M(n)}{D}$. 
The walk following Procedure 3 with the spans between measurements satisfying {\bf Assumption} is 
equivalent to $\DTQWPM{n}$. 
The walk treated in Ref. \cite{SCSK} is a special case of $\DTQWPM{n}$ because 
the walk is a convolution of the independent and \textit{identically} distributed (i.i.d.) 
sequence in Ref. \cite{SCSK}. 
In a general definition of the DTQW with position measurements, however, 
the position state of the DTQW alone is measured at the decided/random times. 
Thereafter, the DTQW is restarted from the measured position remaining the coin state 
by the next measurement time. 
Thus the initial coin state at each measurement time is varied in general 
and depends on the measured position. 
The walk treated here, $\DTQWPM{n}$, is an estimation of the DTQW with PM in that 
reprepared initial coin state is varied but independence of the measurement position. 
The following theorem gives an estimation of the DTQW with the position measurements through 
the weak convergence of the scaled $\DTQWPM{n}$ as the convolution of 
the independent and \textit{not} identically distributed sequence of 
$\{\PPM{j}{D}\}_{{\sf j}=1}^{\infty}$. 
\noindent\\
\begin{theorem}\label{main}
Let $\Theta(n)=\sqrt{\sum_{j=1}^{M(n)}d_j(n)^2}$ and $\sigma^2(a)=1-\sqrt{1-|a|^2}$. 
Then we have as $n\to\infty$, 
\begin{equation}
	\frac{\DTQWPM{n}-E[\DTQWPM{n}]}{\Theta(n)} \Rightarrow N(0,\sigma^2(a)), 
\end{equation}
where $N(a,b)$ is the normal standarddistribution with mean $a$ and variance $b$. 
\end{theorem}
\noindent\\
\quad In heuristic arguments, 
Theorem \ref{main} evaluates the time scaling order $\Theta(n)$ for 
DTQW with position measurements per each time 
with probability $p\sim 1/n^\beta$ $(0<\beta<1)$ 
in the following meaning. 
The range between measurements, $D_j$, is given by the geometric distribution 
with the success probability $p$, that is, $P(D_j=d)=(1-p)^{d-1}p$. 
If we evaluate $D_j$ as its average $d(n)\equiv E[D_j]=1/p\sim n^\beta$, then $M(n)\sim n^{1-\beta}$, 
which is satisfying {\bf Assumption}. 
Thus $\Theta(n)\sim \sqrt{n^{1+\beta}}$. 
Here, for arbitrary functions $f$ and $g$, $f(x)\sim g(x)$ means $\lim_{x\to\infty}f(x)/g(x)=1$. 
This argument is used in Ref. \cite{Turkey} to show dynamics of the
survival probability for the multi-particle DTQW on the ring with the
trap cites. 
As a consequence, Theorem \ref{main} reduces to the result in Ref. \cite{SCSK}: 
\noindent \\
\begin{corollary}[Ref. \cite{SCSK}]\label{PRA}
Assume $\bs{\varphi}_j=\bs{\varphi}_0\equiv q_R\bs{e}_R+q_L\bs{e}_L$ satisfying the symmetric condition (\ref{symmetric}) 
$(j\in\{1,2,\dots\})$. 
Let $d_j(n)\sim d(n)\equiv n^\beta$ with $0\leq\beta\leq 1$ $(j\in\{1,2,\dots\})$. 
So we have as $n\to\infty$, 
\begin{equation}
\frac{\DTQWPM{n}}{\sqrt{n^{1+\beta}}}\Rightarrow 
\begin{cases} 
N(0,1) & \text{: $\beta=0$, } \\ 
N(0,\sigma^2(a))  & \text{: $0<\beta<1$, } \\
K(|a|) & \text{: $\beta= 1$.}
\end{cases}
\end{equation}
\end{corollary}
\noindent
\quad Remark that from Theorem \ref{main}, as long as both $d(n)$ and $M(n)$ are infinite for $n \to \infty$, that is 
$0 <\beta <1$, 
$\DTQWPM{n}$ converges to $N(0,\sigma(a)^2)$ in distribution with the power of the time variable $d(n)\sqrt{M(n)}$.  
At $\beta=0$ (resp. $\beta=1$), $d(n)$ (resp. $M(n)$) is finite in the limit of large $n$. 
The DTQW with PPM always has the discontinuity of the above limit theorem at $\beta=0$ and $\beta=1$. 

As a preparation for the proof of Theorem \ref{main}, we give the useful lemma as follows.  
The proof is given at the end of this section. 
\noindent \\
\begin{lemma}\label{useful}
Let $\PPM{j}{D}$ be a DTQW with the quantum coin $H$ and with an initial coin state 
$\boldsymbol{\varphi}_j$. Then we have 
\begin{equation}
E \left[ e^{i\xi (\PPM{j}{D}-E[\PPM{j}{D}])/\Theta(n)} \right] \sim e^{-\frac{\xi^2}{2}\left(\frac{\sigma(a)d_j(n)}{\Theta(n)}\right)^2}, 
\end{equation}
where an explicit expression for the average $E[\PPM{j}{D}]$ is described in~\cite[Proposition 2]{KonnoQL0}. 
\end{lemma}
Then by using the above lemma, we can give the proof of Theorem \ref{main} in the following. \\
\noindent \\
{\bf Proof. } 
[Proof of Theorem \ref{main}] 
Since $\{\PPM{j}{D}\}_{{\sf j}=1}^{M(n)}$ is an independent sequence, 
the characteristic function for $(\DTQWPM{n}-E[\DTQWPM{n}])/\Theta(n)$ can be written as a product of 
$E[e^{i\xi (\PPM{j}{D}-E[\PPM{j}{D}])/\Theta(n)}]$ $(j\in\{1,2,\dots\})$. 
Thus from Lemma \ref{useful} and the definition of $\Theta(n)$, we can give 
the asymptotic expression for $E[e^{i\xi (\DTQWPM{n}-E[\DTQWPM{n}])/\Theta(n)}]$ 
as follows. 
\begin{align}
E \left[ e^{i\xi (\DTQWPM{n}-E[\DTQWPM{n}])/\Theta(n)} \right] 
	&= \prod_{{\sf{j}}=1}^{M(n)} E[e^{i\xi (\PPM{j}{D}-E[\PPM{j}{D}])/\Theta(n)}]  \notag \\
	&\sim \exp\left[ -\frac{\xi^2}{2} \frac{\sigma^2(a)\sum_{j=1}^M (d_j^{(n)})^2}{\Theta(n)^2} \right] \sim e^{-\xi^2\sigma^2(a)/2}. 
\end{align}
Then we have the desired conclusion. \;\;\;$\Box$\\ 

\noindent Finally, we give the proof of Lemma \ref{useful}. \\

\noindent{\bf Proof. }
[Proof of Lemma \ref{useful}] 
Let us omit the suffix of $d_j(n)$ as $d(n)$. 
We consider the DTQW $\PPM{j}{D}$ at time $d(n)$ with the initial coin state $\bs{\varphi}_0$, 
where $d(n)$ satisfies \AS (\ref{as1}) and (\ref{as2}). 
The spatial Fourier transform for $\bs{\Psi}_{d(n)}(n)$ is described by 
$\widehat{\bs{\Psi}}_{d(n)}(k)\equiv \widehat{H}^{d(n)}(k)\bs{\varphi}_0$, where 
$\widehat{H}(k)\equiv (e^{ik}\bs{e}_R\bs{e}_R^{\dagger}+e^{-ik}\bs{e}_L\bs{e}_L^\dagger)H$. 
According to \cite[Equation (7)]{SK}, the characteristic function for $\PPM{j}{D}$ can be expressed as 
\begin{equation}\label{Sp}
E[e^{i\xi \PPM{j}{D}}]=\int_{0}^{2\pi} \Lambda (k,\xi) \frac{dk}{2\pi}, 
\end{equation}
where $\Lambda (k,\xi)=\langle \widehat{\bs{\Psi}}_{d(t)}(k),\widehat{\bs{\Psi}}_{d(t)}(k+\xi) \rangle$. 
Here, $\langle \bs{u},\bs{v} \rangle$ means an inner product between two vectors $\bs{u}$ and $\bs{v}$. 
From now on, we evaluate $E[e^{i\xi \PPM{j}{D}/\Theta(t)}]$. Note that $\PPM{j}{D}$ is defined at time $d(t)$ from the definition.
Let the eigenvalue and the corresponding eigenvector of $\widehat{H}(k)$ be 
denoted as $e^{i\phi_l(k)}$ and $\bs{v}_l(k)$, respectively ($\l\in\{\pm\}$). 
By using $e^{i\phi_l(k)}$ and $\bs{v}_l(k)$, $\Lambda(k,\xi)$ can be decomposed as 
\begin{multline}\label{decom}
\Lambda (k,\xi) = \sum_{l,m\in\{\pm\}} \exp\left[{id(t)\left\{ \phi_l\left(k+\xi\right)-\phi_m(k) \right\}}\right] \\ 
\times \langle \bs{\varphi}_0, \bs{v}_m(k)\rangle \langle \bs{v}_m(k), \bs{v}_l(k+\xi)\rangle \langle \bs{v}_l(k+\xi), \bs{\varphi}_0\rangle. 
\end{multline}
By replacing $\xi$ with $\xi/\Theta(n)$,  
the term corresponding to the eigenvalues and eigenvectors in Eq. (\ref{decom}) is evaluated as 
\begin{multline}
\exp\left[{id(n)\left\{ \phi_l\left(k+\xi/\Theta(n)\right)-\phi_m(k) \right\}}\right] \\
=e^{id(n)\{ \phi_l(k)-\phi_m(k) \}}\times \exp\left[{i\xi \frac{d(n)}{\Theta(n)} h_l(k)+O\left(\frac{d(n)}{\Theta^2(n)}\right)}\right], 
\end{multline}
\begin{multline}
\langle \bs{\varphi}_0, \bs{v}_m(k)\rangle \langle \bs{v}_m(k), \bs{v}_l(k+\xi/\Theta(n))\rangle \langle \bs{v}_l(k+\xi/\Theta(n)), \bs{\varphi}_0\rangle \\
=\delta_{lm}p_l(k)+\frac{\xi}{\Theta(n)}\langle \bs{\varphi}_0,\bs{v}_m(k)\rangle 
\left\langle \bs{v}_m(k), \frac{\partial}{\partial k} \pi_l(k) \bs{\varphi}_0 \right\rangle +O(\Theta^{-2}(n)),
\end{multline}
where $h_l(k)\equiv \partial \phi_l(k)/\partial k$, $p_l(k)\equiv |\langle \bs{\varphi}_0, \bs{v}_l(k)\rangle|^2$, and 
$\pi_l(k) \equiv \bs{v}_l(k)\;\bs{v}_l(k)^{\dagger}$. 
Here $f(n)=O(g(n))$ means that there exists $0\leq c<\infty$ such that 
$\lim_{n\to\infty}f(n)/g(n)=c$. 
So by the orthonormality of eigenvectors and defining $h(k)\equiv h_+(k)=-h_-(k)$, we obtain 
\begin{equation}
\Lambda (k,\xi/\Theta(n)) = 1-\frac{\xi^2}{2}\left(\frac{d(n)}{\Theta(n)}h(k)\right)^2 
+i\frac{\xi}{\Theta(n)} \mu_{d(n)}(k)+o(\eta(n)), 
\end{equation}
where $f(n)=o(g(n))$ means that $\lim_{n\to\infty}f(n)/g(n)=0$. 
Here, $\mu_{m}(k)\equiv \langle \widehat{\Psi}_{m}(k),D_k \widehat{\Psi}_{m}(k)\rangle$ 
with $D_k=-i\partial/\partial k$ and 
$\eta(n)=d^2(n)/\Theta^2(n)$ (if $\Theta(n)/d^2(n)=o(1)$), $=1/\Theta(n)$ (otherwise). 
Since it is known that (see more details in Eq. (16) of Ref. \cite{Grimmett}, for example. ) 
\begin{equation}
	E[(\PPM{j}{D})^r]=\int_{0}^{2\pi}\langle \widehat{\Psi}_{d(n)}(k), D_k^r \widehat{\Psi}_{d(n)}(k)\rangle\frac{dk}{2\pi}\;\;\;(r\in\{1,2,\dots\}),  
\end{equation}
one obtains 
\begin{equation}
	\int_{0}^{2\pi}\mu_{d(n)}(k) \frac{dk}{2\pi} = E[\PPM{j}{D}]. 
\end{equation}
We can compute the explicit expression for $\int_0^{2\pi}h^2(k)dk/2\pi$ as 
$\sigma^2(a)=1-\sqrt{1-|a|^2}$. 
We have 
\begin{equation}\label{ev}
\int_{0}^{2\pi}\Lambda(k,\xi/\Theta(n))\frac{dk}{2\pi}
	=1+i\frac{\xi}{\Theta(n)}E[\PPM{j}{D}]-\frac{\xi^2}{2}\left(\frac{d(n)}{\Theta(n)}\right)^2 \sigma^2(a) +o \left( \frac{1}{\Theta(n)} \right).
\end{equation}
By multiplying $e^{-i\xi E[Y_{j}^{(D)}]/\Theta(n)}$ to both sides of Eq. (\ref{ev}), we have 
\begin{equation}
E[e^{i\xi (\PPM{j}{D}-E[\PPM{j}{D}])/\Theta(n)}] \sim e^{-\frac{\xi^2}{2}(d(n)/\Theta(n))^2\sigma^2(a)}. 
\end{equation}
Then we obtain the desired conclusion. $\Box$

\subsection{CTQW with position measurements} \label{sec3:2}
Let $\PPM{j}{C}$ be the CTQW at time $d_j(t)$ $(j=1,2,\dots )$ defined in Eq. (\ref{defConti}) with parameter $\gamma$. 
We consider a convolution of an independent sequence $\{\PPM{j}{C}\}_{{\sf j}=1}^{M(t)}$ denoted as 
$\CTQWPM{t} \equiv \PPM{1}{C} + \PPM{2}{C} + \cdots + \PPM{M(t)}{C}$. 
We obtain the following theorem for the continuous-time case in analogy to Theorem \ref{main} in the discrete-time case. 
\noindent \\
\begin{theorem}\label{contiThm}
Let $\Theta^{(C)}(t)=\sqrt{\sum_{j=1}^{M(t)}d_j^2(t)}$. 
We have as $t\to\infty$, 
\begin{equation}
\frac{\CTQWPM{t}}{\Theta(t)} \Rightarrow N(0,|\gamma|^2), 
\end{equation} 
where $\gamma\in\mathbb{C}$ is the parameter of the CTQW defined in Eq. (\ref{defConti}). 
\end{theorem}
\noindent \\
{\bf Proof.} 
{Let the spatial Fourier transform $\widehat{\Psi}_s^{(C)}(k)\equiv \sum_{x\in\mathbb{Z}}\Psi_s^{(C)}(x)e^{ikx}$. 
From Eq. (\ref{defConti}), the solution has 
\begin{equation}
\widehat{\Psi}_s^{(C)}(k)=e^{i|\gamma|s\cos (k+\mathrm{arg}(\gamma))}, 
\end{equation}
where $\mathrm{arg}(z)$ is the argument of the complex number $z$. 
Therefore we have 
\begin{align}
E[e^{i\xi \PPM{j}{C}/\Theta(t)}]
	&=\int_{0}^{2\pi} \langle \widehat{\Psi}_{d_j(t)}^{(C)}(k), \widehat{\Psi}_{d_j(t)}^{(C)}(k+\xi/\Theta^{(C)}(t))\rangle \frac{dk}{2\pi} \notag \\
	&= 1-\frac{\xi^2}{2}\left( \frac{d_j(t)}{\Theta^{(C)}(t)} \right)^2 +o\left(\frac{d_j(t)}{\Theta^{(C)}(t)}\right)^2.
\end{align}
From the independence of $\{\PPM{j}{C}\}_{{\sf j}=1}^{\infty}$, we have 
\begin{align}
E[e^{i\xi \CTQWPM{t}/\Theta^{(C)}(t)}] &= \prod_{j=1}^{M(t)}E[e^{i\xi \PPM{j}{C}/\Theta^{(C)}(t)}] \notag \\
&\sim \exp\left[ -\frac{\xi^2}{2} \frac{\sum_{j=1}^{M(t)} (d_j(t))^2}{\Theta^{(C)}(t)^2} \right]
\to e^{-\xi^2/2} 
\end{align}
as $t\to\infty$}
\begin{flushright} $\square$ \end{flushright}
It is emphasized that the CTQW with position measurements is nothing but expressed by $\CTQWPM{t}$ without approximation 
in contrast with the previous discrete-time case, since the CTQW is described only by the position-state space.
\begin{corollary}
Let $d_j(t)\sim t^\beta$ with $0\leq\beta\leq 1$ $(j\in\{1,2,\dots\})$. 
So we obtain as $t\to\infty$, 
\begin{equation}
\frac{\CTQWPM{t}}{\sqrt{t^{1+\beta}}}\Rightarrow 
\begin{cases} 
N(0,|\gamma|^2)  & : 0\leq\beta<1, \\
Z(\gamma) & : \beta= 1.
\end{cases}
\end{equation}
\end{corollary}
There is no discontinuous point at $\beta=0$ unlike the DTQW with PPM. 
According to Theorem 2, the CTQW with PPM always has the 
discontinuity of the above limit distribution at $\beta=1$ (corresponding to ``finite measurements'') 
and continuity at $\beta=0$ (corresponding to ``finite span between measurements'').
\section{Crossover from DTQW to CTQW} \label{sec4}
To give some insights into 
the similarity of the results on the position measurements for discrete-time and continuous-time cases, 
in this section, we introduce a final-time-dependent walks which are modified walks initiated by 
Strauch~\cite{Strauch1}. 
Let $n$ be the final time, that is, a particle keeps walking until the final time comes. 
At first, we show a construction of the CTRW from final-time-dependent DTRWs. 
Secondly, we give CTQWs in some limit of the FTD-DTQW 
which is a quantum analogue of the final time dependent RW. 
Throughout this paper, 
we assume the parameter $r(n)>0$ with $r(n)\to 0$ as $n\to \infty$ for the final-time dependent DTRW and DTQW. 
\subsection{RW case} \label{4:1}
We consider the final-time-dependent RW on $\mathbb{Z}$, 
where the ``final time'' means the time that a particle stops the walk.  
The average of the waiting time of particle movement is $1/r(n)$, 
where $r(n)$ is the probability that the particle moves by the final time $n$. 
The number of particle movements in the walk by the final time $n$ is evaluated as $r(n)n$. 
A particle spends her most time being lazy since the rate of movements $r(n)$ tends to $0$ as $n\to\infty$. 
This is called a lazy RW. In the following, we will show that the lazy RW with the number of particle's 
movements $nr(n)$ can be taken as the CTRW with the final time $t$ for sufficiently large $n$.

Let $\LA{m}$ be the lazy RW at time $m \in \{0,1,2,\dots,n\}$ defined by 
\begin{equation}\label{lazyeq}
	p_{m}(x)=(1-r(n))p_{m-1}(x)+\frac{r(n)}{2} \bigg\{p_{m-1}(x-1)+p_{m-1}(x+1)\bigg\},\;\;p_0(x)=\delta_0(x), 
\end{equation}
where $p_m(x) \equiv P(\LA{m}=x)$. 
This means that a particle stays at the same place with the probability $1-r(n)$, and 
the particle jumps left or right with probability $1/2$ when the moving opportunity comes with probability $r(n)$. 
Put $\bs{\mu}_m={}^T[\dots,p_m(-1),p_m(0),p_m(1),\dots]$. Equation (\ref{lazyeq}) is equivalent to 
\begin{equation}
	\bs{\mu}_m=\left\{\ID+r(n)(A/2-\ID)\right\} \bs{\mu}_{m-1},\;\;\bs{\mu}_0=\bs{\delta}_0, 
\end{equation}
where $A\bs{\delta}_x=\bs{\delta}_{x+1}+\bs{\delta}_{x-1}$ and $\ID$ is the identity operator. 
Therefore we have under the 
assumption of $nr^2(n)\to 0$ as $n\to \infty$, 
\begin{equation}
 \bs{\mu}_n \sim e^{nr(n)(A/2-\ID)}\bs{\mu}_0.
\end{equation}
Replace the particle movements $nr(n)$ with a continuous parameter $t$. 
We can express $p_n (x)\sim m_t (x)$ for sufficiently large $n$, where 
$m_s(x)$ satisfies 
\begin{equation}\label{ContiRW}
	\frac{\partial }{\partial s}m_{s}(x)=\frac{1}{2}\left\{m_s(x+1)+m_s(x-1)\right\}-m_s(x), 
	\;\;m_0(x)=\delta_0(x)\;\;\;\;(s\leq t). 
\end{equation}
This differential equation corresponds to the CTRW with the final time $t$. 
We have the central limit theorem by 
using the Fourier transform in the following: 
if $nr(n) \to \infty$, as $n \to \infty$, then 
\begin{equation} \label{LazyCentral}
	\frac{\LA{n}}{\sqrt{nr(n)}} \Rightarrow N(0,1)\;\;(n\to \infty). 
\end{equation}
It is remarked that Eq. (\ref{LazyCentral}) can be shown without $nr(n) \to 0$ as $n \to \infty$
In particular, in the case of $r(n) \sim r/n^\alpha$ with 
$0 < r <1$ and $0 \leq \alpha$, we obtain a crossover from the diffusive to the localized spreading: 
as $n\to \infty$, if $0 \leq \alpha <1$, then 
\begin{equation}\label{DRtoCR}
	\frac{\LA{n}}{\sqrt{n^{1-\alpha}}} \Rightarrow N(0,r)
\end{equation}
and if $\alpha=1$, then 
\begin{equation} \label{DRtoCR2}
	P(\LA{n}=x) \sim e^{-r}I_x(r), 
\end{equation}
where $I_\nu(z)$ is the modified Bessel function of order $\nu$. As for the modified 
Bessel function, see Ref. \cite{Watson}. 
Note that Eq.(\ref{DRtoCR2}) comes from the correspondence between $nr(n)$ and $t$, and 
the Fourier transform for Eq. (\ref{ContiRW}): 
for large $n$, 
\begin{equation}
p_n(x) \sim m_t(x)=e^{-t}I_x(t). 
\end{equation}
If $\alpha>1$, then we can easily see that $P(\LA{n}=x) \to \delta_0(x)$. 
The limit theorem for the lazy RW will be used in Sect.~\ref{sec5}. 
\subsection{QW case} \label{sec4:2}
In this subsection, we will consider a quantum analogue of the above method of continuum approximation to the lazy RW. 
The final-time-dependent quantum coin is defined by 
\begin{equation}\label{Qcoin}
	H_n=\begin{bmatrix} \sqrt{r(n)} & \sqrt{1-r(n)}  \\  \sqrt{1-r(n)} & -\sqrt{r(n)}  \end{bmatrix}. 
\end{equation}
The quantum coin is a quantum analogue to the final-time-dependent stochastic coin of the correlated RW~\cite{KonnoCor}. 
We give a relation between the correlated RW and the FTD-DTQW, 
and the limit theorems for the correlated RW in Appendix A. 
According to Ref. \cite{Strauch1}, 
the absolute values of diagonal parts of the quantum coin are sufficiently small and independent of the final time $n$. 
According to Ref.~\cite{Romanelli}, the quantum coin is changed at each time, 
and its diagonal parts at time $m(<n)$ are given in proportion to $1/m^\alpha$ ($0\leq \alpha\leq 1$). 
On the other hand, in our model, all elements of the quantum coin depend on the final time $n$. 
The above quantum coin (\ref{Qcoin}) shows that a particle moves the same direction of the previous step with 
the probability amplitude $\sqrt{r(n)}$ (left case) and $-\sqrt{r(n)}$ (right case), 
and the opposite one with $\sqrt{1-r(n)}$ (left and right cases).
This is called an FTD-DTQW. 
We give the following lemma which shows that 
the FTD-DTQW is expressed as a linear combination of some CTQWs for sufficiently large $n$. 
The following lemma is consistent to Refs.~\cite{Strauch1,Romanelli} except for the time scaling. 
Let $n$ be the final time for the FTD-DTQW and $t$ be the final time for the corresponding CTQWs. 
We define a quantum analogue of the waiting time of a quantum particle movement by $n/t$. 
\noindent \\
\begin{lemma}\label{DtoC}
Let $\bs{\Psi}_n^{(F)}(x)$ be the coin state of the FTD-DTQW at time $n$ and position $x$. 
Put $t=n\sqrt{r(n)}$ with $nr(n)= o(1)$ for large $n$. 
$\bs{\Psi}_n^{(F)}(x)$ is asymptotically expressed by 
\begin{equation}
\bs{\Psi}_n^{(F)}(x) \sim \frac{1}{2}\left( \bs{\Psi}_t^{(+)}(x)+(-1)^n\bs{\Psi}_t^{(-)}(x) \right)
\end{equation}
with $t=n\sqrt{r(n)}$, where $\bs{\Psi}_s^{(\pm)}(x)$ $(0<s<t)$ satisfies the following Schr\"{o}dinger equation:
\begin{align}
-i\frac{\partial}{\partial s}\Psi_s^{(\pm,J)}(x) &=\pm\frac{1}{2}\left(i\Psi_s^{(\pm,J)}(x-1)
-i\Psi_s^{(\pm,J)}(x+1)\right),\;\; (J\in \{L,R\}) \label{qwsch} \\
\Psi_0^{(\pm,L)}(x) &= q_L\delta_0(x)\pm q_R\delta_1(x),\; \;\Psi_0^{(\pm,R)}(x)=q_R\delta_0(x)\pm q_L\delta_{-1}(x), \label{qwsch1}
\end{align} 
where $\Psi_s^{(\pm,J)}(x)=\langle \bs{e}_{J},\bs{\Psi}_s^{(\pm)}(x) \rangle$, $(J\in\{R,L\})$. 
\end{lemma}
It is noted that Eq. (\ref{qwsch}) can be taken as the CTQW with $\gamma = \pm i$ and the modified initial condition 
Eq. (\ref{qwsch1}). \\
\noindent \\
{\bf Proof. }
By the spatial Fourier transform for Eq. (\ref{Qcoin}), we obtain 
\begin{align}
	\widehat{H}_n^2(k)= \ID - 2i\sqrt{r(n)}\sin k V_{\sigma_x}(k)+O(r(n)), 
\end{align}
with $V_{\sigma_x}(k)=\left(e^{-ik}\bs{e}_L\bs{e}_L^\dagger+e^{ik}\bs{e}_R\bs{e}_R^{\dagger}\right){\sigma_x}$, 
where ${\sigma_x}$ is $x$ comportment of the Pauli matrix. 
When $|\mathrm{det}(\sqrt{r(n)}\sin k V_{\sigma_x}(k))|<1$, so 
$\log(\widehat{H}_n^2(k))=-2i\sqrt{r(n)}\sin k V_{\sigma_x}(k)+O(r(n))$, 
where $\mathrm{det}(A)$ is the determinant of $A$. 
Therefore we have under the assumption of $nr(n)\to 0$ as $n\to \infty$, 
\begin{equation} \label{DtoCMaster}
\widehat{H}^{n}(k)=(V_{\sigma_x}(k))^{n}e^{-in\sqrt{r(n)}\sin k V_{\sigma_x}(k)+O(nr(n))}
\sim (V_{\sigma_x}(k))^{n} e^{-in\sqrt{r(n)}\sin k V_{\sigma_x}(k)}. 
\end{equation}
Because of $(\ID \pm V_{\sigma_x}(k))e^{-is\sin k V_{\sigma_x}(k)}=e^{\mp is\sin k}(\ID \pm V_{\sigma_x}(k))$ 
for any real number $s$, 
we obtain 
\begin{equation}\label{applox1}
	\widehat{H}^{n}(k)\sim \frac{1}{2}\left( e^{-in\sqrt{r(n)}\sin k }(\ID+V_{\sigma_x}(k))+(-1)^{n} e^{in\sqrt{r(n)}\sin k}(\ID-V_{\sigma_x}(k)) \right). 
\end{equation}
To see a relation between the FTD-DTQW and the CTQW, 
we define $\widehat{\bs{\Psi}}_s^{(\pm)}(k)\equiv e^{\mp is\sin k}\widehat{\bs{\Psi}}^{(\pm)}_0(k)$ with 
$\widehat{\bs{\Psi}}^{(\pm)}_0(k)\equiv (\ID \pm V_{\sigma_x}(k))\bs{\varphi}_0$. 
It is noted that 
$\widehat{\bs{\Psi}}_{n}(k)=\left(\widehat{\bs{\Psi}}_{t}^{(+)}(k)+(-1)^{n}\widehat{\bs{\Psi}}_{t}^{(-)}(k)\right)/2$, 
with $t=n\sqrt{r(n)}$ and $\widehat{\bs{\Psi}}_{s}^{(\pm)}(k)$ obeys 
\begin{equation}
\pm i \frac{d}{ds}\widehat{\bs{\Psi}}_s^{(\pm)}(k)=\sin k \widehat{\bs{\Psi}}_s^{(\pm)}(k).
\end{equation}
\begin{flushright}$\Box$\end{flushright}
By the inverse Fourier transform for Eq. (\ref{applox1}) and the definition of the Bessel function, 
we get the following theorem. 
\noindent \\
\begin{theorem}\label{kamakura}
Let $\FTD{n}$ be the FTD-DTQW at time $n$. 
Put $t=n\sqrt{r(n)}$. 
When the initial coin state is $\bs{\varphi}_0=q_L\bs{e}_L+q_R\bs{e}_R$ with $|q_L|^2+|q_R|^2=1$, then we have 
\begin{equation}
	P(\FTD{n}=x)\sim \frac{1+(-1)^{n+x}}{2}\mathcal{J}(x;t), 
\end{equation}
where 
\begin{equation}
	\mathcal{J}(x;t)=\left\{1-(\overline{q_R}q_L+q_R\overline{q_L})\frac{2x}{t}\right\}J_x^2(t)+|q_L|^2J_{x-1}^2(t)+|q_R|^2J_{x+1}^2(t).
\end{equation}
Here $J_\nu(z)$ is the Bessel function of the first kind of order $\nu$. 
As for the Bessel function, see Ref. \cite{Watson}. 
\end{theorem}
\noindent \\
The following theorem shows that 
the weak convergence theorem of the FTD-DTQW also holds 
without the assumption of $nr(n)\to 0$ as $n \to \infty$. 
\noindent \\
\begin{theorem}\label{COQCD}
Let the initial coin state be $\bs{\varphi}_0=q_L\bs{e}_L+q_R\bs{e}_R$ with $|q_L|^2+|q_R|^2=1$. 
Assume $n \sqrt{r(n)} \to \infty$ as $n \to \infty$. 
Then we have as $n \to \infty$, 
\begin{equation}\label{WCDtoC}
	\frac{\FTD{n}}{n\sqrt{r(n)}}\Rightarrow A^{\bs{\varphi}_0}(1), 
\end{equation}
where $A^{\bs{\varphi}_0}(r)$ has the following density:
\begin{equation}
\rho(x;r)=\left\{1-(q_L\overline{q_R}+\overline{q_L}q_R)x/r\right\}\frac{\IND_{(-|r|,|r|)}(x)}{\pi \sqrt{r^2-x^2}}. 
\end{equation} 
\end{theorem}
{\bf Proof.} 
{We show that Eq. (\ref{WCDtoC}) holds without assumption $nr(n)\to 0$ as $n\to \infty$. 
The Fourier transform for the quantum coin $H_n$ (\ref{Qcoin}) is described as 
\begin{equation}
	\widehat{H}_n(k)=
	\begin{bmatrix} e^{-ik}\sqrt{r(n)} & e^{-ik}\sqrt{1-r(n)} \\ e^{ik}\sqrt{1-r(n)} & -e^{ik}\sqrt{r(n)}  \end{bmatrix}. 
\end{equation}
Let the eigenvalue and corresponding eigenvector of $\widehat{H}_n(k)$ 
be $e^{i\theta^{(\pm)}_n(k)}$ and $\bs{v}^{(\pm)}_n(k)$. 
Then we obtain 
\begin{align}
\cos \theta^{(\pm)}_n(k) &= \pm \sqrt{1-r(n)\sin ^2 k},\;\;\sin \theta^{(\pm)}_n(k)=-\sqrt{r(n)}\sin k, \\
\pi^{(\pm)}_n(k) \equiv \bs{v}^{(\pm)}_n(k) {\bs{v}^{(\pm)}_n(k)}^\dagger &= \frac{1}{2}\left(I\pm V_{\sigma_x} (k) \right) + O \left( \sqrt{r(n)} \right). 
\end{align}
Let $\FTD{s}$ be the FTD-DTQW at time $s$ with the initial coin state 
$\bs{\varphi}_0=q_L\bs{e}_L+q_R\bs{e}_R$. 
It is noted that 
\begin{equation}
	\theta_n^{(\pm)}(k+\xi/t)-\theta_n^{(\pm)}(k)=\mp \xi \frac{\sqrt{r(n)}}{t}\cos k
        +O\left(\frac{(\sqrt{r(n)})^3}{t}\right),
\end{equation}
where $t=n\sqrt{r(n)}$. We have 
\begin{align}
E[e^{i\xi \FTD{n}/t}] 
	&=\int_{0}^{2\pi} \langle \widehat{H}^n(k)\bs{\varphi}_0, \widehat{H}^n(k+\xi/t) \bs{\varphi}_0\rangle \frac{dk}{2\pi} \notag \\
	&=\int_{0}^{2\pi} 
        	\sum_{j\in \{0,1\}} e^{-(-1)^j i\xi\cos k+O\left(\sqrt{r(n)}\right)} \notag \\
        &\;\;\;\;\;\;\;\;\;\times \frac{1}{2}\langle \bs{\varphi}_0,\left(I+(-1)^j V_{\sigma_x}(k)\right) \bs{\varphi}_0 \rangle\frac{dk}{2\pi}+O(\sqrt{r(n)}) \notag \\
        &\to \int_{-\infty}^{\infty}e^{i\xi x} \bigg\{1-(q_L \overline{q_R}+\overline{q_L}q_R) x \bigg\}\frac{\IND_{(-1,1)}(x)}{\pi\sqrt{1-x^2}}dk
        \;\; (n\to \infty),
\end{align}
where $t=n\sqrt{r(n)}$. Then we get the desired conclusion. }
\begin{flushright}$\square$\end{flushright}

From Theorems \ref{kamakura} and \ref{COQCD}, we obtain the following limit theorem 
which shows a crossover from the localized to the ballistic spreading. 
\noindent \\
\begin{corollary}\label{DQtoCQ}
If $\sqrt{r(n)}=r/n^{\alpha}$ with $0<r<1$, then as $n\to \infty$, 
\begin{equation}
\frac{\FTD{n}}{n^{1-\alpha}} \Rightarrow 
	\begin{cases}
        	K(r) & {\rm if} \ \alpha = 0, \\ 
        	A^{\bs{\varphi}_0} (r) & {\rm if} \ 0< \alpha < 1, 
	\end{cases} 
\end{equation}
and if $\alpha=1$, then 
\begin{equation}
	P( \FTD{n} =x ) \sim \frac{1+(-1)^{n+x}}{2} \mathcal{J}(x;r). 
\end{equation}
and if $\alpha>1$, then $P(\FTD{n} =x) \to \delta_0(x)$ as $n\to\infty$. 
\end{corollary}
{\bf Proof.} 
{Since $\alpha=0$ case corresponds to the DTQW, $\FTD{n}/n \Rightarrow K(r)$ as $n \to \infty$. 
The $\sqrt{r(n)}=n^\alpha$ with $0 < \alpha < 1$ satisfies the condition $n \sqrt{r(n)} \to 0$ as 
$n \to \infty$. Therefore it follows from Theorem \ref{COQCD} that 
$\FTD{n}/n^{1-\alpha} \Rightarrow A^{\bs{\varphi}_0}$ with $0<\alpha<1$. 
The result on $\alpha \geq 1$ case derives from Theorem \ref{kamakura}. }
\begin{flushright}$\square$\end{flushright}

\section{From DTQW to CTQW with PPM} \label{sec5}
\begin{figure}[htbp]
\begin{center}
	\includegraphics[width=150mm]{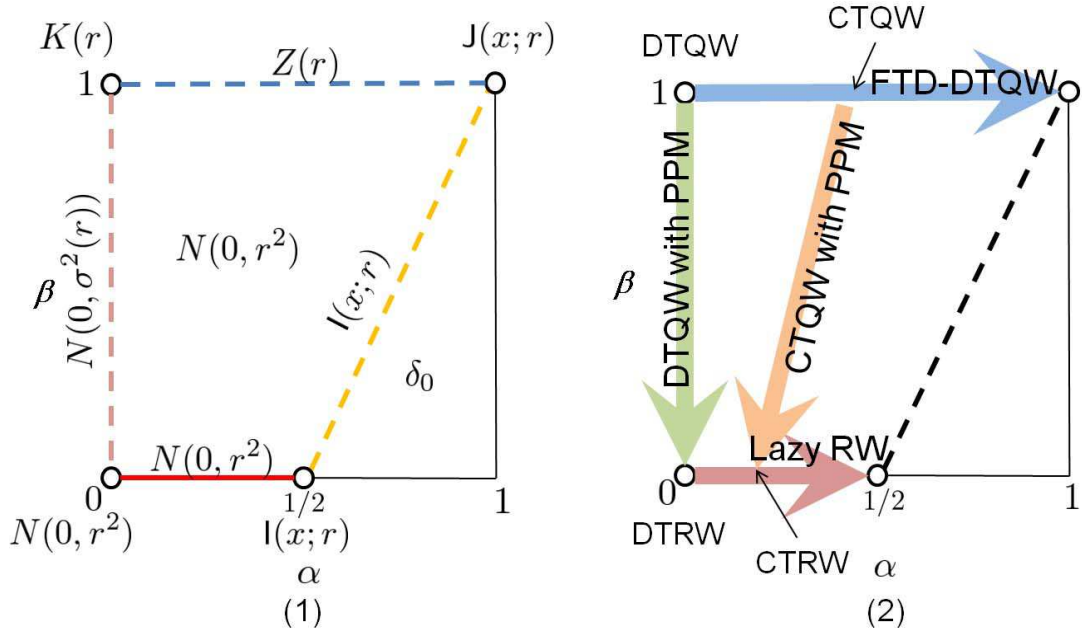}
\end{center}
\caption{(1) The $\alpha$-$\beta$ diagram with respect to limit distributions in Theorem \ref{MainResult} is shown. 
The corners, $(\alpha,\beta)=(0,1)$ and $(\alpha,\beta)=(0,0)$, correspond to the DTQW and DTRW, respectively. 
If $(\alpha,\beta)\in S$, then $W_n^{(n)}/\sqrt{2^{1-\beta}n^\beta-2\alpha+1}$ converges weakly to 
$N(0,r^2)$ as $n\to\infty$. (2) The boundary of $S$ is mapped to the DTQW with PPM ($\alpha=0$), 
the FTD-DTQW ($\beta=1$) and the lazy RW ($\beta=0$). 
If $0<\alpha<1$, the CTQW (resp. CTRW) corresponds to the FTD-DTQW (resp. lazy RW) with the continuous 
final time $t\sim n^{1-\alpha}$ (resp. $s\sim n^{1-2\alpha}$). 
Therefore the CTQW is mapped to the point $(\alpha,1)$ while the CTRW is $(\alpha/2,0)$ with $0<\alpha<1$. 
Thus the CTQW with PPM can be seen as a line which connects the point $(\alpha,1)$ to $(\alpha/2,0)$ 
with $0<\alpha<1$. }
\label{fig:one}
\end{figure}

We give a relation between the DTQW with PPM and the corresponding CTQW. 
In this section, for simplicity, 
let $d_j(n)=d(n)$ for all $j\in\{1,2,\dots\}$ and $\bs{\varphi}_j=\bs{e}_L$ with probability $1/2$, 
$=\bs{e}_R$ with probability $1/2$. 
We consider a hybrid type walk, that is, 
the position of the FTD-DTQW is measured per $d(n)$th step analogues to Ref.~\cite{SCSK}. 
To give the following relation to the lazy RW discussed in the previous section, here we assume that $d(n)$ is even. 
\noindent \\
\begin{remark}\label{d(n)}
Assume that the final time $n$ is even.  
Let $\QW{n}$ be the FTD-DTQW with PPM at time $n$. 
The following two cases 
correspond to the lazy RW and the FTD-DTQW, respectively: 
\begin{enumerate}
\item $d(n)=2$ ($\Leftrightarrow M(n)=n/2$) case: 
This case does not satisfy \As (\ref{as2}) in Sect.~\ref{sec3}. 
This walk corresponds to the FTD-DTQW with position measurements per two steps (not each step). 
Let $\LA{m}$ be the lazy RW at time $m$ defined in Eq. (\ref{lazyeq}). If $x$ is even, then we have 
$P(\QW{n}=x)=P(\LA{n/2}=x/2)$. \label{dn1}
\item $d(n)=n$ ($\Leftrightarrow M(n)=1$) case: 
This case dose not satisfy \As (\ref{as3}) in Sect.~\ref{sec3}. 
This walk corresponds to the FTD-DTQW without position measurements by the final time, \textit{i.e.}, 
$P(\QW{n}=x)=P(\FTD{n}=x)$. 
\end{enumerate}
\end{remark}
\noindent \\
The next theorem is our main result which gives limit theorems for the FTD-DTQW with PPM when 
both the span between measurements $d(n)$ and the number of measurements $M(s)$ go to infinity 
as $n\to\infty$, simultaneously in contrast with Remark \ref{d(n)}. 
\noindent \\
\begin{theorem}\label{MainResult}
Let $S=\{(\alpha,\beta)\in[0,1]^2: 2\alpha<1+\beta,0<\alpha<1, 0\leq\beta<1\}\cup\{(0,0)\}$ 
and $S^{\prime} = \{(\alpha,\beta)\in[0,1]^2: 2\alpha=1+\beta,0\leq\alpha\leq 1, 0\leq\beta\leq 1  \}$. 
When we take $d(n)\sim 2^{1-\beta}n^\beta$ and $r(n)\sim r^2/n^{2\alpha}$ with $0\leq \alpha,\beta\leq 1$ and $0<r<1$, 
we obtain an $\alpha$-$\beta$ phase diagram with 
respect to limit distribution (see Fig.~\ref{fig:one} (1)):  
\begin{equation}
\frac{\QW{n}}{\sqrt{2^{1-\beta}n^{\beta-2\alpha+1}}} \Rightarrow 
	\begin{cases}
        	K(r) & : ( \alpha, \beta )=(0,1), \\
                Z(r) & : \beta =1, 0 < \alpha <1,  \\
                N(0,\sigma^2(r)) & : \alpha =0, 0< \beta <1, \\
                N(0,r) & : ( \alpha, \beta ) \in S. 
        \end{cases}
\end{equation}
If $( \alpha, \beta ) \in \{( \alpha , \beta ) \in [0,1]^2 : 2 \alpha \geq 1 + \beta \}$, then 
\begin{equation}
\lim_{n\to\infty}P(\QW{n}=x)=
	\begin{cases} 
            {\sf{I}}(x;r) & : (\alpha , \beta ) \in S^{\prime}, 0 \leq \beta < 1, \\
            {\sf{J}}(x;r) & : (\alpha , \beta ) \in S^{\prime}, \beta =1, \\
            \delta_0(x) & : {\rm otherwise},
        \end{cases}
\end{equation}
where
${\sf{J}}(x;r)=\IND_{\{x:even\}}(x)\times \left\{J^2_x(r)+\left( J^2_{x-1}(r)+J^2_{x+1}(r) \right)/2\right\}$ 
and ${\sf{I}}(x;r)=\IND_{\{x:even\}}(x)\times \left\{ e^{-r^2/2}I_{x/2}\left(r^2/2\right) \right\}$. 
\end{theorem}
The limit distributions in the case of $0\leq \alpha, \beta \leq 1$ for the FTD-DTQW with PPM can be illustrated 
in Fig. \ref{fig:one} (1) as the phase diagram.
Theorem~\ref{MainResult} shows that in the limit of $n \to \infty$, the FTD-DTQW with PPM parametrized by 
$\alpha$ and $\beta$ corresponds to the walks which have been treated in this paper like Fig. \ref{fig:one} (2). \\
The following lemma is essential to the proof of Theorem \ref{MainResult}. 
The proof of the lemma is given at the end of this section. 
\noindent \\
\begin{lemma}\label{hybrid}
Define $\widetilde{\Theta}(n)=d(n)\sqrt{r(n)M(n)}$. We assume that $d(n)$ is even and satisfies \AS (\ref{as1}) - (\ref{as3}). 
Then we have 
\begin{enumerate}
\item When $\widetilde{\Theta}(n)\to \infty$ as $n\to\infty$, 
\begin{equation}
	\frac{\QW{n}}{\widetilde{\Theta}(n)} \Rightarrow N(0,1). 
\end{equation} \label{sa1}
\item When $\widetilde{\Theta}(n)\to c_0<\infty$ as $n\to\infty$,  
if $x$ is odd, then we have $P(\QW{n}=x)=0$, and if $x$ is even, then we have 
\begin{equation}
\lim_{n\to\infty}P(\QW{n}=x)
	= \IND_{\{x:even\}}(x)\times 
\begin{cases} 
	e^{-c_0^2/2}I_{x/2}\left(c_0^2/2\right) & : c_0>0, \\ 
	\delta_0(x) & : c_0=0. 
\end{cases}
\end{equation} \label{sa2}
\end{enumerate}
\end{lemma}
\noindent\\

\begin{remark}
We show how the CTQW with PPM can be seen from Lemma \ref{hybrid} in the following. 
Let $\CTQWPM{t}$ be the CTQW with PPM at time $t$ defined in Eq. (\ref{defConti}). 
Define the span between measurements of the CTQW by $d(t)$. 
Then we have from Theorem \ref{contiThm}, as $t\to \infty$, 
\begin{equation}
	\frac{\CTQWPM{t}}{d(t)\sqrt{M(t)}} \Rightarrow N(0,1) \label{ContiC}. 
\end{equation}
As we have already seen in Sect. 3, the quantum analogue of the waiting time is evaluated as 
$1/\sqrt{r(n)}$. Then we can evaluate $t \sim n\sqrt{r(n)}$, $d(t)\sim \sqrt{r(n)}d(n)$, and 
$M(n)=n/d(n) \sim n\sqrt{r(n)}/d(t) \sim t/d(t)=M(t)$. Therefore combining the above estimations with 
Lemma \ref{hybrid}, we obtain Eq. (\ref{ContiC}). 
\end{remark}
\noindent\\
{\bf Proof. }[Proof of Theorem \ref{MainResult}]  
The DTQW with PPM corresponds to $\alpha=0$, $0\leq \beta\leq 1$ with the quantum coin 
$r\bs{e}_L\bs{e}_L^\dagger+\sqrt{1-r^2}\bs{e}_R\bs{e}_L^\dagger+r\bs{e}_L\bs{e}_R^{\dagger}-\sqrt{1-r^2}\bs{e}_R\bs{e}_R^{\dagger}$. 
So Corollary \ref{PRA} and Remark \ref{d(n)} (\ref{sa1}) give the desired conclusion in the case of $\alpha=0$ 
and $0\leq \beta\leq 1$. 
In a similar way, both ``$\beta=0$, $0\leq \alpha\leq 1/2$'' and ``$\beta=1$, $0\leq \alpha\leq 1$'' cases 
correspond to the lazy RW and the FTD-DTQW, respectively. 
Thus the results on these two cases follow from Eqs. (\ref{DRtoCR}), (\ref{DRtoCR2}) and Corollary \ref{DQtoCQ}. 
Finally, Lemma \ref{hybrid} implies the other cases, i.e., $0<\alpha,\beta<1$. $\Box$ \\

\noindent We now give the proof of Lemma \ref{hybrid}. \\

\noindent{\bf Proof. } [Proof of Lemma \ref{hybrid}] 
Let $\{\PPM{j}{F}\}_{{\sf j}=1}^{M(n)}$ be an i.i.d. sequence of the FTD-DTQW at time $d(n)$ with 
$\QW{n} = \PPM{1}{F} + \PPM{2}{F} + \cdots + \PPM{M(n)}{F}$.
\begin{enumerate}
\item $\widetilde{\Theta}(n)\to \infty$ case: 
The Fourier transform for the quantum coin $H_n$ (\ref{Qcoin}) is expressed as 
\begin{multline}\label{applox}
\widehat{H}^{d(n)}(k)= 
	e^{-id(n)\arcsin\left(\sqrt{r(n)}\sin k\right)}\pi_n^{(+)}(k) \\
	+(-1)^{d(n)} e^{id(n)\arcsin \left(\sqrt{r(n)}\sin k\right)} \pi_n^{(-)}(k). 
\end{multline}
Note that 
\begin{equation} 
\pi_n^{(l)}(k+\xi/\widetilde{\Theta}(n))\cdot {\pi_n^{(m)}(k)}^{\dagger} 
	= \delta_{lm}\pi_n^{(l)}(k)+O(1/\widetilde{\Theta}(n)) \;\;(l,m\in\{\pm\}), 
\end{equation}
\begin{multline}
	\arcsin\left(\sqrt{r(n)}\sin (k+\xi/\widetilde{\Theta}(n))\right)-\arcsin\left(\sqrt{r(n)}\sin k\right) \\ 
	= \frac{\xi}{d(n)\sqrt{M(n)}} \cos k+O\left(\frac{r(n)}{d(n)\sqrt{M(n)}}\right). 
\end{multline}
From Eq. (\ref{applox}), we have 
\begin{align}
E[e^{i\xi Y_j^{(n)}/\widetilde{\Theta}(n)}] &= 
	\int_0^{2\pi} \langle \widehat{H}^{d(n)}(k)\bs{\varphi}_0, \widehat{H}^{d(n)}(k+\xi/\widetilde{\Theta}(n))\bs{\varphi}_0\rangle \frac{dk}{2\pi} \notag \\
 	&= \sum_{j\in\{\pm\}} \int_0^{2\pi} e^{ij\xi\left\{ \cos k+O(r(n)) \right\}/\sqrt{M(n)}}
        			\times  \frac{1}{2}\mathrm{Tr}(\pi_n^{(j)}(k))\frac{dk}{2\pi} \notag \\
	&\sim 1-\frac{\xi^2}{2}\times \frac{1}{M(n)}+o\left(\frac{1}{M(n)}\right). 
\end{align}
Since $\{\PPM{j}{F}\}_{{\sf j}=1}^{M(n)}$ is an i.i.d. sequence, we have as $n \to \infty$, 
\begin{equation}
	E[e^{i\xi \QW{t}/g(n)}] \to e^{-\xi^2/2}. 
\end{equation}
\item $\widetilde{\Theta}(n)\to c_0<\infty$ case: 
Combining $d(n)\sqrt{r(n)}=o(1)$ with Theorem \ref{kamakura}, we obtain 
\begin{equation}
	P(\PPM{1}{F}=x) \sim \frac{1+(-1)^{d(n)+x}}{2}{\sf{J}}(x;d(n)\sqrt{r(n)}), 
\end{equation}
where ${\sf{J}}(x;s)=J^2_x(s)+\left( J^2_{x-1}(s)+J^2_{x+1}(s) \right)/2$. \\
Note that $J_x^2(s)\sim \delta_0(x)+\left\{\delta_0(x-1)/2+\delta_0(x+1)/2-\delta_0(x)\right\}s^2$ 
for sufficiently small $s$. 
Thus the characteristic function of $\PPM{1}{F}$ is evaluated as 
\begin{equation}
	E[e^{i\xi \PPM{1}{F}}]
	=\sum_{x:even}J^2_x(s)e^{i\xi x} + \cos \xi \sum_{x:odd} J^2_x (s) e^{i\xi x} = 1-d^2(n) r(n) \left\{\sin^2\xi + o(1) \right\}.
\end{equation}
So we have as $n\to\infty$, 
\begin{equation}
	E[e^{i\xi \QW{n}}] \sim \left\{ 1-d^{2}(n)r(n)\sin^2\xi \right\}^{M(n)} \to 
	\begin{cases} 
        	e^{-c_0^2\sin^2\xi} & : c_0>0, \\
		1 & : c_0=0. 
	\end{cases}
\end{equation}
It is noted that because of 
$\sum_{x \in \mathbb{Z}} e^{-r} I_x(r) e^{2i\xi x} = e^{-2r\sin^2\xi}$ and Remark \ref{d(n)} (\ref{sa1}), 
the value $e^{-2 r \sin^2 \xi}$ is the limit of the characteristic function of the scaled lazy RW:  
$2 \times \LA{n/2}$ with $nr(n) \to r$ as $n \to \infty$. 
Then we complete the proof. $\Box$
\end{enumerate}
\section{Summary} \label{sec6}
We have analyzed long-time behaviors for DTQWs with position measurements from the view point of the weak convergence theorem. 
In the situation that both the span of the position measurements and its number are simultaneously infinite 
as the final time goes to infinity; $n \to \infty$, we have shown a crossover from ballistic spreading to diffusive spreading of the particle. 
Physically speaking, we have analytically seen the long-time behavior 
in the decoherence model of the discrete- and 
continuous-time QWs. Our main result is summarized in Theorem~\ref{MainResult} and is illustrated in Fig.~\ref{fig:one}. 
To show this crossover, we have given the estimation of the limit theorems for the DTQW with position measurements 
per each time with probability $p \sim 1/n^\beta$ (Theorem \ref{main}). 
This result generalizes our previous study~\cite{SCSK}. 
Also, we have obtained the similar result on the corresponding CTQW (Theorem~\ref{contiThm}). 
Furthermore, we have introduced 
a new class of the QW, FTD-DTQW, inspired by Refs.~\cite{Strauch1,Romanelli}. We have given the estimation of the 
limit theorem for the FTD-DTQW. This means that the DTQW and CTQW are connected for the sufficiently long-time behavior 
(Corollary~\ref{DQtoCQ}). We have presented the relationship between the FTD-DTQW with position measurements per two steps 
and the lazy RW (Remark~\ref{d(n)} (\ref{dn1})). These analytical results are our contributions in this paper.

\par
\
\par\noindent
\noindent
{\bf Acknowledgments.}
One of the authors (Y.S.) acknowledges encouragement from Seth Lloyd. 
N.K. is supported by the Grant-in-Aid for Scientific Research (C) (No. 21540118). 
Y.S. is supported by JSPS Research Fellowships for Young Scientists (No. 21008624).
\par
\
\par

\begin{small}
\bibliographystyle{jplain}

\end{small}

\noindent\\
\noindent\\
\noindent\\

\noindent {\large{\bf Appendix A}} {Correlated RW} \\
\noindent\\ 
\noindent
Because of the No-Go Lemma~\cite{Meyer}, we cannot construct 
the quantum analogue of the lazy RW directly. 
The final-time-dependent quantum coin is a quantum analogue to the stochastic coin $C_n$ which 
determines the behavior of the correlated RW: 
\begin{equation}\label{Scoin}
	C_n= \begin{bmatrix} r(n) & 1-r(n) \\  1-r(n) & r(n) \end{bmatrix}. 
\end{equation}
The coin of the correlated RW depends on the previous time: 
a particle moves to the opposite direction of the previous time with probability $1-r(n)$ 
and moves to the same direction with probability $r(n)$. 
According to the assumption for $r(n)$; $r(n) \to 0$ as $n \to \infty$,
a particle tends to walk \textit{zigzag}. 
The number of opportunities in the walk by the final time 
that a particle moves the same direction of the previous time is evaluated as $nr(n)$. 
Let $\CO{m}$ be the correlated RW at time $m$. 
The distribution is determined by $P(\CO{m}=x)= \langle \bs{e},\bs{p}_m(x)\rangle$, 
where $\bs{e}=\bs{e}_L+\bs{e}_R$. 
Here 
$\bs{p}_m(x)=u\bs{e}_L+v\bs{e}_R$ with $0\leq u,v\leq 1$ and $0\leq u+v\leq 1$ 
satisfies the following relations: 
\begin{equation}\label{CorRWDef}
\bs{p}_m(x)=\widetilde{Q}_n\bs{p}_{m-1}(x-1)+\widetilde{P}_n\bs{p}_{m-1}(x+1),
\;\;\bs{p}_0(x)=\delta_0(x)\bs{\phi}_0, 
\end{equation}
where $\widetilde{P}_n=\bs{e}_L\bs{e}_L^\dagger C_n$, $\widetilde{Q}_n=\bs{e}_R\bs{e}_R^{\dagger}C_n$, and 
$\bs{\phi}_0={}^T[p_L,p_R]\in [0,1]^2$ with $p_L+p_R=1$. 
The detailed asymptotic estimation of $\bs{p}_n(x)$ for sufficiently large $n$ 
is almost similar to the FTD-DTQW case, so we give the results without the proof. 
Let $t=nr(n)$ under the assumption $nr^2(n)=o(1)$, and the initial state is 
$\bs{p}_0(x)=\delta_0(x)\left(p_L\bs{e}_L+p_R\bs{e}_R\right)$. 
Then we have for sufficiently large $n$, 
\begin{equation}
\bs{p}_n(x)\sim \frac{1}{2}\left\{\bs{p}^{(+)}_t(x)+(-1)^t\bs{p}^{(-)}_t(x)\right\}, 
\end{equation}
where $\bs{p}^{(\pm)}_t(x)=p^{(\pm,R)}\bs{e}_R + p^{(\pm,L)}\bs{e}_L$ satisfies 
\begin{align}
\frac{\partial }{\partial t}p_{t}^{(\pm,J)}(x)
	&=\pm \frac{1}{2}\left(p_{t}^{(\pm,J)}(x+1)
        +p_{t}^{(\pm,J)}(x-1)\right)-p_{t}^{(\pm,J)}(x) \;\; (J\in \{L,R\}), \\
p_0^{(\pm,L)}(x) 
	&= p_L\delta_0(x)\pm p_R\delta_1(x),\;\;p_0^{(\pm,R)}(x)=p_R\delta_0(x)\pm p_L\delta_{-1}(x).       
\end{align}
Here, the ``$+$'' part corresponds to the CTRW given by Eq. (\ref{ContiRW}).
We obtain the asymptotic behavior of $P(\CO{n}=x)$ as follows.  
Let $nr^2(n)\to 0$ as $n\to \infty$. 
For sufficiently large $n$, we have 
\begin{equation}
	P(\CO{n}=x)\sim \frac{1+(-1)^{n+x}}{2}e^{-t}
		\bigg\{ p_LI_{x-1}(t)+I_x(t)+p_RI_{x+1}(t) \bigg\}, 
\end{equation}
where $t=nr(n)$. Analogous to the lazy RW, we also get the following weak convergence theorem: 
for any initial state $\bs{p}_0(x)=\delta_0(x)\bs{\phi}_0$, we have with $nr(n)\to\infty$, 
\begin{equation}
	\frac{\CO{n}}{\sqrt{nr(n)}} \Rightarrow N(0,1) \;\; (n\to \infty). 
\end{equation}

\end{document}